\documentclass{article}
\usepackage{spconf,amsmath,graphicx}
\usepackage{hyperref}
\usepackage{cite}
\usepackage{float}
\usepackage{tikz}
\usetikzlibrary{shapes,arrows, snakes}

\tikzstyle{specialblock} = [draw, ultra thick, fill=blue!20, rectangle, 
    minimum height=3em, minimum width=4em]
\tikzstyle{block} = [draw, fill=lightgray, rectangle, 
    minimum height=3em, minimum width=4em]
\tikzstyle{sum} = [draw, fill=white, circle, node distance=1cm]
\tikzstyle{prod}   = [circle, minimum width=8pt, draw, inner sep=0pt, path picture={\draw (path picture bounding box.south east) -- (path picture bounding box.north west) (path picture bounding box.south west) -- (path picture bounding box.north east);}]
\tikzstyle{sumt}   = [circle, minimum width=8pt, draw, inner sep=0pt, path picture={\draw (path picture bounding box.east) -- (path picture bounding box.west) (path picture bounding box.south) -- (path picture bounding box.north);}]
\tikzstyle{input} = [coordinate]
\tikzstyle{output} = [coordinate]
\tikzstyle{pinstyle} = [pin edge={to-,thin,black}]
\tikzset{
tmp/.style  = {coordinate}, 
dot/.style = {circle, minimum size=#1,
              inner sep=0pt, outer sep=0pt},
dot/.default = 6pt 
}
\usepackage{amssymb}
\usepackage{amsmath}

\usepackage{mathtools}

\usepackage{capt-of}
\usepackage{url}

\title{REAL-M: Towards Speech Separation on Real Mixtures}
\name{Cem Subakan$^{1, 2}$, Mirco Ravanelli$^2$, Samuele Cornell$^3$, Fran\c{c}ois Grondin$^1$}
\address{$^1$Universit\'{e} de Sherbrooke, Canada,  
$^2$Mila-Quebec AI Institute, Canada,\\
$^3$Università Politecnica delle Marche, Italy} 

\begin{document}
\ninept
\maketitle
%

\begin{abstract}
In recent years, deep learning based source separation has achieved impressive results. Most studies, however, still evaluate separation models on synthetic datasets, while the performance of state-of-the-art techniques on in-the-wild speech data remains an open question. This paper contributes to fill this gap in two ways. First, we release the REAL-M dataset, a crowd-sourced corpus of real-life mixtures. Secondly, we address the problem of performance evaluation of real-life mixtures, where the ground truth is not available. We bypass this issue by carefully designing a  blind  Scale-Invariant Signal-to-Noise Ratio (SI-SNR) neural estimator. Through a user study, we show that our estimator reliably evaluates the separation performance on real mixtures. The performance predictions of the SI-SNR estimator indeed correlate well with human opinions. Moreover, we observe that the performance trends predicted by our estimator on the REAL-M dataset closely follow those achieved on synthetic benchmarks when evaluating popular speech separation models. 
\end{abstract}
\section{Introduction}
\label{sec:intro}

Source separation techniques have evolved quickly in the last few years and recently reached impressive performance levels. SepFormer \cite{subakan2020attention}, Wavesplit \cite{zeghidour2020wavesplit}, and DualPath RNNs \cite{luo2020dualpath}, for instance, achieve more than 20 dB improvement in the Scale Invariant Signal-to-Noise Ratio (SI-SNR) \cite{le2019sdr}.
The vast majority of studies in the field employ synthetic datasets for both training and evaluation purposes \cite{ luo2018convtasnet, chen2020dualpath, wang2018deep, nachmani2020voice, tzinis2020sudo, hershey2015deep}.
This practice is largely accepted by the community and offers undeniable advantages, as simulated data are easily created from clean recordings. The widely used WSJ0-2/3Mix datasets \cite{hershey2015deep}, which are generally considered the de-facto standard benchmark,
consist of signals that are artificially produced by mixing speech recordings from the Wall Street Journal (WSJ) corpus \cite{WSJ}. The original utterances are recorded with high-quality microphones in controlled environments where there is no noise and reverberation.
Another noteworthy advantage is that artificial mixtures can be synthesized on the fly using dynamic mixing \cite{zeghidour2020wavesplit, tzinis2019twostep}. This feature is extremely valuable when training neural models, as we can augment the dataset and improve the separation performance significantly.

Simulated mixtures are commonly used for evaluation purposes as well. Despite being widely adopted, we think that this practice is potentially misleading. It might bias the community towards systems that work well in laboratory conditions and fail in real-life scenarios. 
Some efforts have been recently devoted to mitigating this lack of realism. 
For instance, WHAM! \cite{wichern2019wham} and WHAMR! \cite{maciejewski2020whamr} datasets introduced environmental noise and reverberation, but still provide simulated data for evaluation that are not entirely representative of the challenges faced with in-the-wild signals.
Evaluating real-life mixtures would arguably be the safest approach to avoid mismatches between laboratory and real-life conditions. The main challenge that prevents it is the lack of ground truth signals (i.e., the clean sources used as targets for training and evaluating speech separation). Specialized hardware and controlled recording conditions can circumvent this issue, but this practice is costly, time-consuming, and impractical to adopt widely.

This paper addresses the problem from a different angle and provides a two-fold contribution. 
First, we release REAL-M, a real-life speech source separation dataset for two-speaker mixtures. We collected this dataset through crowdsourcing by asking contributors to read a predefined set of sentences simultaneously. The mixtures are recorded in different acoustic environments using a wide variety of recording devices such as laptops and smartphones, thus reflecting more closely potential application scenarios. REAL-M currently contains more than 1400 speech mixtures from 50 native and non-native English speakers, for a total of almost three hours of real-world speech mixtures with their corresponding transcriptions. 

As a second contribution, we carefully design and study a neural network that blindly estimates the quality of a signal processed by a speech separation model. The idea of automatically assessing speech quality using a neural network has been recently explored in the literature.
In \cite{yu2021metricnet, reddy2021dnsmos, manocha2021noresqa,fu2019metricgan, fu2021metricgan+}, for instance, 
authors designed neural networks that estimate non-differentiable speech quality metrics (e.g., PESQ) for improving speech enhancement. These works showcase that estimating speech quality without having access to the ground truth is feasible. We here inherit the same philosophy in the context of speech separation. In particular, we employ a convolutional model to predict the SI-SNR from the audio mixture and separated signals.

To the best of our knowledge, this paper is the first work showing that neural metric learning is reliable enough for accurately assessing speech separation on real mixtures.
It turned out that the estimates of the SI-SNR performance on the REAL-M dataset
correlate reasonably well with human opinion scores. We also evaluated popular speech separation models using both the standard simulated data and REAL-M. The predictions of the SI-SNR estimator on REAL-M follow a trend that closely resembles the performance observed on simulated data, giving further confirmation on the effectiveness of the proposed approach.

With REAL-M, we also release \href{https://github.com/speechbrain/speechbrain/tree/develop/recipes/REAL-M/sisnr-estimation}{the training script of the SI-SNR estimator} within the SpeechBrain \cite{speechbrain}, and the pre-trained SI-SNR estimator on the \href{https://huggingface.co/speechbrain/REAL-M-sisnr-estimator}{Huggingface}.
The dataset is available on our website\footnote{\url{https://sourceseparationresearch.com/static/REAL-M-v0.1.0.tar.gz}}.

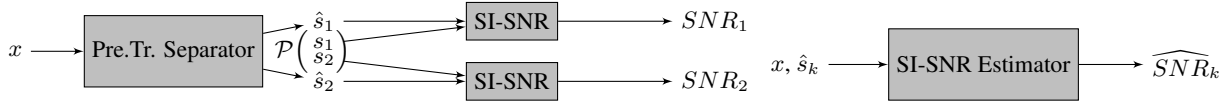
\begin{figure*}[t]

\centering
    \newcommand{\sepfig}{0.5}

  \begin{tikzpicture}[auto, node distance=2.5cm,>=latex']
        \node [draw=none, fill=none] (x) {$x$};
        \node [block, right of=x, fill, xshift=-0.4cm] (1) {Pre.Tr. Separator};  
        \node [draw=none, fill=none, right of=1, yshift=0.4cm, xshift=-0.5cm] (shat1) {$\hat{s}_1$};
        \node [draw=none, fill=none, right of=1, yshift=-0.4cm, xshift=-0.5cm] (shat2) {$\hat{s}_2$};
        \node [draw=none, fill=none, right of=1, xshift=-0.5cm] (s1s2) {};
        \node [draw=none, fill=none, right of=1, xshift=-0.65cm] (s1s2p) {$\mathcal{P}\Big (  \; \; \; \; \Big) $};
        \node [draw=none, fill=none, right of=1, xshift=-0.5cm, yshift=0.1cm] (s1) {$s_1$};
        \node [draw=none, fill=none, right of=1, xshift=-0.5cm, yshift=-0.1cm] (s2) {$s_2$};
        \node [block, right of=s1s2, fill, xshift=0cm, yshift=0.4cm, minimum size=0.5cm] (sisnr1) {SI-SNR} ;  
        
        \node [block, right of=s1s2, fill, xshift=0cm, yshift=-0.4cm, minimum size=0.5cm] (sisnr2) {SI-SNR} ;  
        \node [draw=none, fill=none, right of=s1s2, yshift=0.4cm, xshift=2.7cm] (snr1) {$SNR_1$};
        \node [draw=none, fill=none, right of=s1s2, yshift=-0.4cm, xshift=2.7cm] (snr2) {$SNR_2$};
        
        \draw [->] (x) -- (1);
        \draw [->] (1) -- (shat1);
        \draw [->] (1) -- (shat2);
        \draw [->] (shat1) -- (sisnr1);
        \draw [->] (shat2) -- (sisnr2);
        \draw [->] (s1) -- (sisnr1);
        \draw [->] (s2) -- (sisnr2);
        
        \draw [->] (sisnr1) -- (snr1);
        \draw [->] (sisnr2) -- (snr2);
  \end{tikzpicture}
  \begin{tikzpicture}[auto, node distance=2.5cm,>=latex']
        \node [draw=none, fill=none] (1) {$x$, $\hat{s}_k$};
        \node [block, right of=1, fill, xshift=0cm] (est) {SI-SNR Estimator};  
        \node [draw=none, fill=none, right of=1, yshift=-0.0cm, xshift=2.7cm] (snrhat2) {$\widehat{SNR}_k$};
        
        \draw [->] (1) -- (est);
        \draw [->] (est) -- (snrhat2);
  \end{tikzpicture}
  \caption{Training the neural SI-SNR estimator. A pretrained separator is used to estimate the sources (left). The SI-SNR is computed using the ground truth signals (center). The SI-SNR Estimator is fed by the mixtures and the estimated sources and predicts the SI-SNRs (right). }
  \label{fig:snrest}

\end{figure*}

\section{Data Collection}


%

The data collection has been conducted as follows.
We showed the participants the text of two sentences randomly sampled from the test set of LibriSpeech \cite{Panayotov_librispeech} dataset. We limited the length of the sentences to be between 5 to 15 words. In total, we presented 569 unique pairs of sentences to the participants, and we collected 1436 mixtures. We asked the contributors to simultaneously read the shown utterances while being  physically in the same room. Beyond that, we also recorded 144 mixtures where one participant was recorded through a videoconferencing software (e.g., Zoom, google meet). This increases the diversity of the dataset and addresses an application scenario that frequently occurs these days. 

Thanks to crowd-sourcing, the acoustic conditions and recording equipment encompass a wide variety of scenarios: mixtures contain varying levels of reverberation and noise as the speech can be either near or far-field.
Moreover, we involved both native and non-native speakers with different accents, including American, British, French, Italian, Persian, Indian, and African. 
The recordings have been performed with a purposely built \href{https://sourceseparationresearch.com}{data collection platform} interfaced with Amazon Mechanical Turk. 


\section{Blind Neural SI-SNR Estimation}
As humans, we can tell if the quality of a speech signal is good or not just by listening to it. By following this reasoning, we describe here how we designed a neural network that estimates the separation performance in terms of SI-SNR without accessing the ground truth signals. 

\subsection{Training} 
The basic components needed to train the neural SI-SNR estimator are a pretrained separation model, a synthetic dataset, and a neural network for SI-SNR estimation. The training pipeline, shown in Fig. \ref{fig:snrest}, comprises of the following steps:

\begin{enumerate}
    \item First, we  process the synthetic mixtures $x$ with a pretrained speech separation model, as reported in the following equation:
    \begin{equation}
    \hat{s}_1, \hat{s}_2 \; = \; \text{PT-S}(x).  \\
    \end{equation}
    This step provides separated signals $\hat{s}_1$ and $\hat{s}_2$ (with different levels of distortions) that the neural estimator will assess. The parameters of the pretrained separator are kept frozen.
    \item Then, we use the ground truth sources $s_k$ and the separated sources $\hat{s}_k$ to compute the oracle SI-SNR values:
    \begin{equation}
    {SNR}_k \; = \; \text{SI-SNR}(s_k, \hat{s}_k), \; k \in \{1, 2\}.      
    \end{equation}
    As shown in Figure \ref{fig:snrest}, the permutation over ground truth sources is resolved before calculating the oracle SI-SNR values. The oracle SI-SNR values $SNR_k$ are used as a target for the SI-SNR estimator. 
    \item We feed the separated signals $\hat{s}_k$ into the neural estimator, which aims to predict the SI-SNR performance:
     \begin{equation}
    \widehat{SNR}_k \; = \; \text{SI-SNR-Estimator}(x, \hat{s}_k),  \; k \in \{1, 2\}      
    \end{equation}   
    
    The model is fed by the mixture signal as well. This addition leads to more accurate predictions, as it provides a more accurate guideline for the neural estimator. 

\item The neural estimator is trained to regress the oracle SI-SNR values, as shown in the following equation:
    \begin{equation}
     \mathcal{L} \; = \; \| SNR_1 - \widehat{SNR}_1 \|_1 + \| SNR_2 - \widehat{SNR}_2 \|_1, 
    \end{equation}
where $SNR_{1,2}$ are the oracle SI-SNRs computed in step 2 and $\widehat{SNR}_{1,2}$ are the estimated SI-SNR values computed at step 3. Training is conducted in a standard way using back-propagation coupled with the ADAM optimizer \cite{kingma2017adam}. More details on the training pipeline can be found on the SI-SNR estimator recipe released in SpeechBrain.
    
\end{enumerate}

\subsubsection{Synthetic Mixture Creation}
The training pipeline involves a synthetic dataset composed of artificial mixtures with their corresponding ground truths. 
In this work, we consider the LibriMix\cite{cosentino2020librimix}, and the WHAMR! \cite{maciejewski2020whamr} datasets simultaneously by randomly choosing mixtures from the two datasets. 

More precisely, we create the mixtures on the fly using dynamic mixing \cite{zeghidour2020wavesplit, tzinis2019twostep}. We randomly sample relative mixing SNRs in the range between 0-5dB. Environmental noise (using noisy sequences from the WHAM!  corpus \cite{wichern2019wham}) and reverberation (using the impulse responses of the WHAMR! \cite{maciejewski2020whamr} dataset) are added as well. The synthetic mixtures are processed by the pretrained separation models described in the following sub-section.

\subsubsection{Pretrained Separation Models}
We consider two different ways of using pretrained separators in the training pipeline:
\begin{itemize}
    \item (\emph{single}). We train the SI-SNR estimator using only one separation model. In our case, we used the \href{ https://huggingface.co/speechbrain/sepformer-whamr}{SepFormer model pretrained on the WHAMR! dataset} from the SpeechBrain Hugging Face repository \cite{subakan2020attention}.
\item (\emph{pool}). We train the SI-SNR estimator using a pool of many source separation models. Namely, we used a mixture of SepFormer \cite{subakan2020attention}, DPRNN \cite{luo2020dualpath} and Convtasnet \cite{luo2018convtasnet}. At training time, one model from the pool is uniformly sampled and applied to the mixture to derive the estimated sources. 
For each model, we use the checkpoints after the first and last epochs. Moreover, we use a checkpoint in the middle of training. In total, we use a combination of 9 pretrained separators. The \textit{pool} approach leads to wide variability in the SI-SNR values. The estimator can observe different speech distortions and artifacts, with benefits on its generalization properties.
\end{itemize}

\subsection{Inference}
At inference time, we can simply feed the separated signal and the corresponding mixture to the neural estimator. The latter will provide an estimate of the SI-SNR performance.
The proposed approach is blind because it does not require accessing the clean ground truth sources at inference time. It also has the advantage of being very straightforward and fast.
\vspace{-0.3cm}
\subsection{Architecture}
The SI-SNR estimator consists of five convolutional layers (with a kernel size of 4, 128 channels, and stride of 1), followed by a statistical pooling layer, and two fully connected layers (with 256 neurons). We use ReLU activations for all the layers.  We limit the estimated SI-SNR values to fall between 0 and 10 dB.
This range covers the typical SI-SNRs observed on reverberant source separation datasets such as WHAMR! \cite{maciejewski2020whamr}.
We also normalize the network output to fall between 0-1, where 0 is associated with 0 dB and 1 is associated with 10 dB. This range compression is performed with a sigmoid applied at the output of the network. The sources estimates and the mixture are normalized to have zero-mean and unit-variance before inputting them to the SI-SNR estimator. These normalization steps are important to ensure an accurate prediction and a fast convergence of the model. 

The estimator employs about 300K trainable parameters. It is thus compact and suitable for fast evaluations. We performed an extensive architecture search to find the adopted hyperparameter setting.

\vspace{-0.3cm}
\section{RESULTS}
In this section, we first assess the reliability of the Neural SI-SNR Estimator on synthetic data (using LibriMix and WHAMR!). Then, we will show the results achieved with the REAL-M dataset. 

\vspace{-0.2cm}
\subsection{Results on LibriMix and WHAMR!}
\label{sec:synthsnr-estimation}
\begin{figure}[h!]
    \centering
    \includegraphics[width=0.18\textwidth]{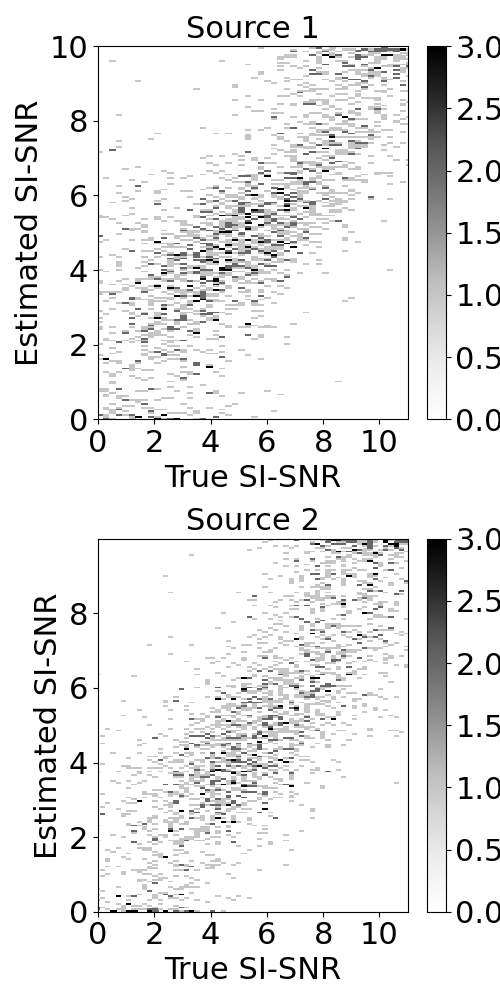}
    \includegraphics[width=0.18\textwidth]{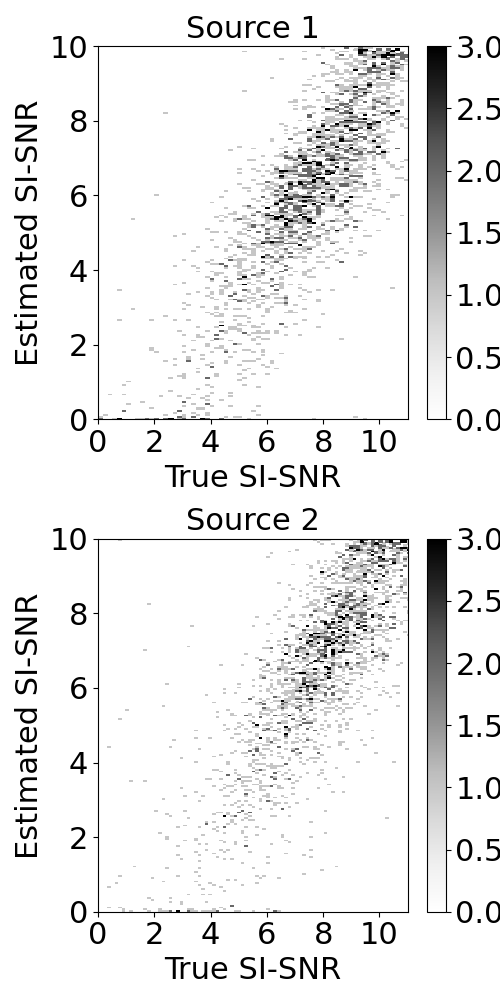}
    \caption{Correlation plots for the SI-SNR estimates vs the ground truth SI-SNR values on the LibriMix dataset (left) and the WHAMR! dataset (right).}
    \label{fig:snrscatter1}
\end{figure}

Figure \ref{fig:snrscatter1} shows the scatter plots of the ground truth SI-SNRs and the estimated ones for the \emph{single} training strategy. In this figure, we evaluate the SI-SNR estimator on the test sets of LibriMix and WHAMR! datasets when processing the mixtures with the SepFormer \cite{subakan2020attention}.

\begin{figure}[h!]
    \centering
    \includegraphics[width=0.18\textwidth]{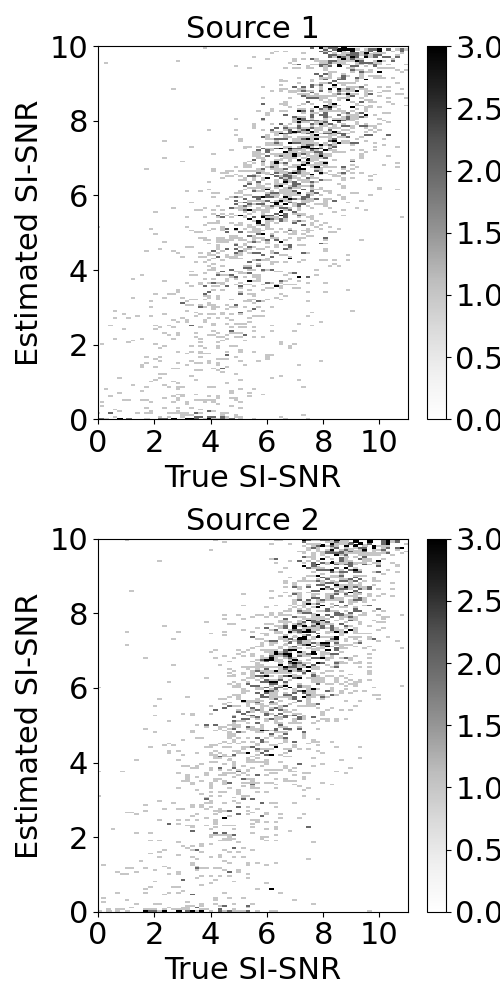}
    \includegraphics[width=0.18\textwidth]{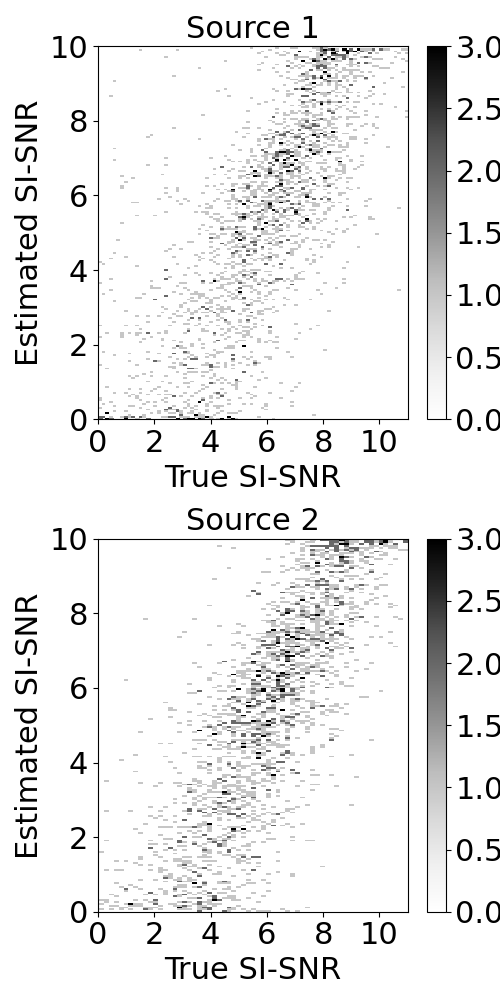}
    \caption{Correlation plots for the SI-SNR estimates vs the ground truth obtained with Dual-Path RNN (left) and Convtasnet (right) on the WHAMR! dataset (mismatched conditions)}
    \label{fig:corr-othermodels}
\end{figure}

Interestingly, we notice a strong correlation. The Pearson coefficient is indeed around 0.8 for both datasets. 
To ascertain whether the SI-SNR estimator works with different separation models, we tested it with mixtures processed by DPRNN and ConvTasNet (CTN). Note that the estimator is trained with the SepFormer only when adopting the \emph{single} training strategy. Figure \ref{fig:corr-othermodels} shows the scatter plots under this mismatched condition.  We can see that, even when tested with different separator models, the SI-SNR estimation correlates well with  the oracle values. The Person correlation is still around 0.8 in all mismatched conditions. 

In Table \ref{table:synthcorrelation} we compare the \emph{single} and \emph{pool} training strategies described before.  As expected, we observe a larger Pearson correlation  with the \emph{pool} strategy,  which confirms the intuition that the SI-SNR estimator robustness is improved when training it with an ensemble of separators. The estimator is trained on a larger variety of distortions and artifacts that improves its robustness and generalization.
For the rest of this work concerning real-world data, we use this SI-SNR estimator. 

The results discussed in this section provide a preliminary indication of the effectiveness of the proposed approach. In the next section, we will extend the analysis to real-life mixtures.

\begin{table}[t!]
 \caption{Pearson Correlation Values (averaged over sources) for Different Separators on LibriMix and WHAMR! datasets, with two different SI-SNR estimators.}
 \label{table:synthcorrelation}
 \centering
\resizebox{8.0cm}{!}{
\begin{tabular}{l|c|c|c|c}
 & \multicolumn{2}{|c|}{SI-SNR-Estimator 1 \emph{(single)}} & \multicolumn{2}{|c}{SI-SNR-Estimator 2} \emph{(pool)}   
  \\ \hline 
\textbf{Model} & \textbf{LibriMix} & \textbf{WHAMR!} & \textbf{LibriMix} & \textbf{WHAMR!} \\
\hline \hline 
SF &  0.80 & 0.81  & 0.82 & 0.87  \\ \hline 
DPRNN & 0.80  & 0.80 & 0.83  & 0.84   \\ \hline 
CTN & 0.81 & 0.79 & 0.85 & 0.86  \\ \hline 
\end{tabular}
}
\end{table}

\begin{figure}[ht]
    \centering
    \includegraphics[width=0.48\textwidth, trim=0.5cm 0cm 0.1cm 0cm, clip]{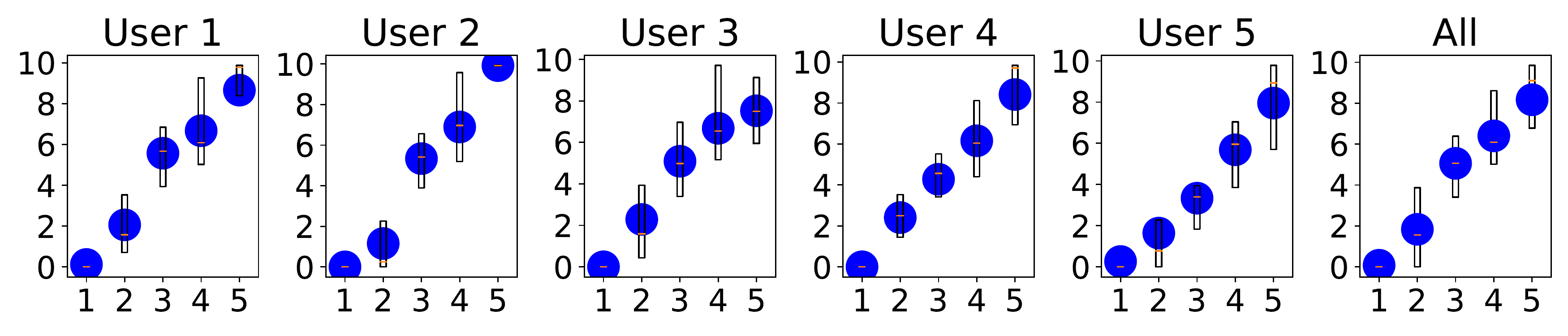}
    
    \vspace{-0.2cm}
    \caption{A user study on estimated SI-SNR vs human opinion. The x-axis represents the participant preference (1 is `bad' quality, while 5 is an `excellent` quality). Y-axis represents the estimated SI-SNR. Blue Circles represent the mean values, and the bars represent the 25th-75th percentile of each distribution.}
    \label{fig:mos}
\end{figure}

\subsection{Results on REAL-M Dataset} 
\subsubsection{Subjective Opinion Scores}
REAL-M is a dataset of real-life mixtures, and therefore no oracle SI-SNR is available.
To validate the predictions of the neural estimator on real-world mixtures, we conducted a user study. We gathered the opinions of participants by asking them to assign a score between 1 (\textit{`bad'} separation) and 5 (\textit{`excellent'} separation) to the estimated sources.

For this study, we selected a subset of 50 mixtures and the corresponding estimated sources obtained with SepFormer trained on WHAMR!. We chose the mixtures to have a uniform distribution of estimated SI-SNR values between 0-10 dB.  
Figure \ref{fig:mos} shows the scores obtained from five participants along with the aggregated one (rightmost plot).
Each subplot shows the distribution of the estimated SI-SNR values for each of the opinion scores.

We observe that, on average,  the estimated SI-SNR values are highly correlated with user opinions. Despite some variability, the average SI-SNR estimations linearly match with the human scores. In particular, the 25th-75th percentile of the distributions (shown with error bars around the mean of each distribution) correlates nicely with the user opinion scores. 
This result suggests that, on average, the neural estimator provides reliable predictions on the REAL-M dataset as well.


\subsubsection{Estimated SI-SNR over training epochs}
\label{sec:snr-epoch}
\vspace{-0.0cm}

\begin{figure}[t!]
    \centering
    \includegraphics[trim=1.0cm 0cm 1cm 0cm , clip, width=0.5\textwidth]{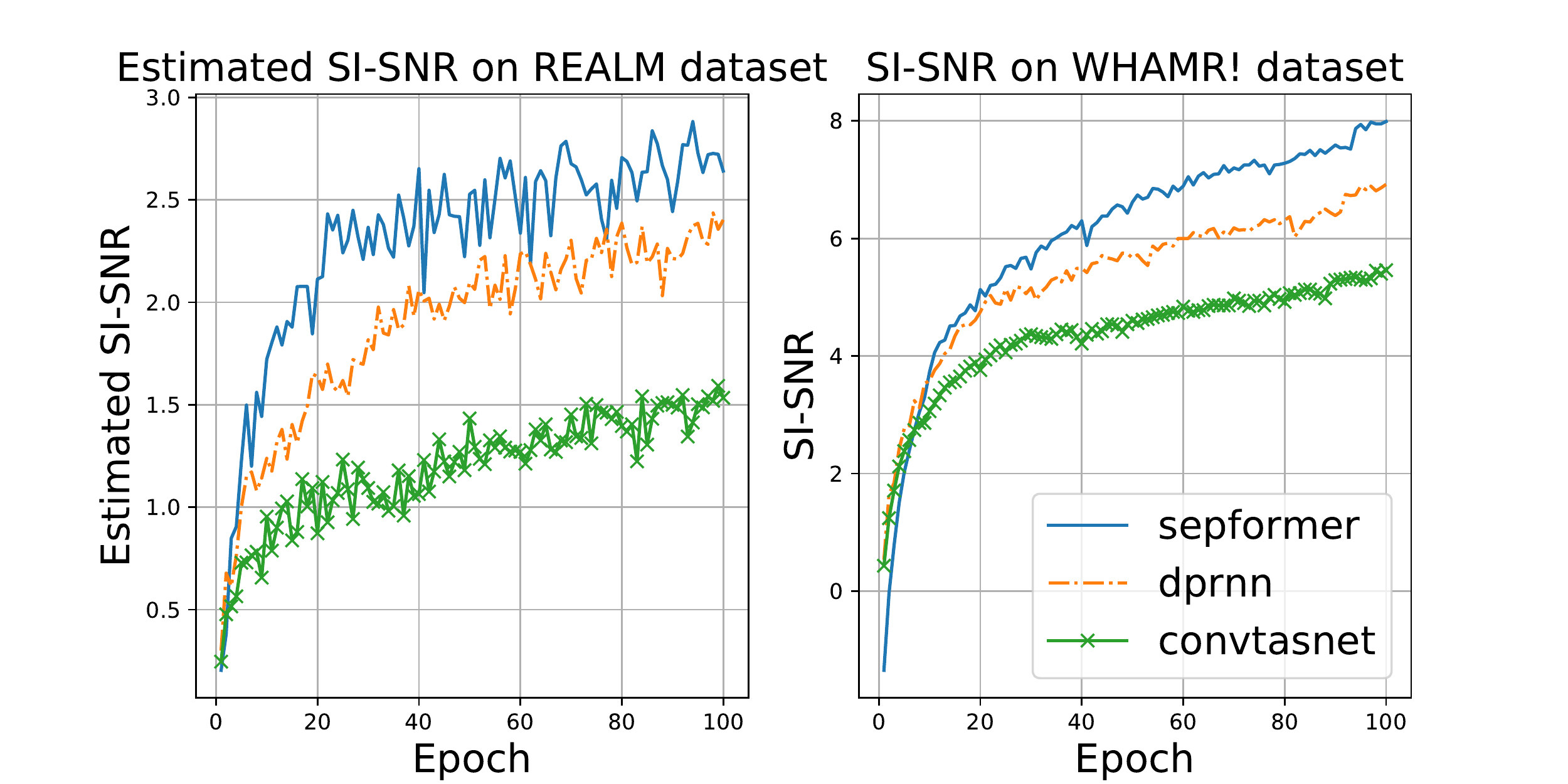}
    \vspace{-0.5cm}
    \caption{Estimated SI-SNR vs training epochs for three popular source-separation models on the REALM dataset. (left) Achieved SI-SNR on the validation set of WHAMR! at the same epoch as left figure (right). }
    \label{fig:snrhatvsepoch}
\end{figure}

We now investigate how the estimated SI-SNR changes over the training epochs. 
Figure \ref{fig:snrhatvsepoch} shows the curves obtained when training the SepFormer \cite{subakan2020attention}, Dual-Path RNN \cite{luo2020dualpath}, and Convtasnet \cite{luo2018convtasnet} from scratch on the WHAMR! dataset.
Figure \ref{fig:snrhatvsepoch} (left) shows the SI-SNR estimated by the proposed network when using REAL-M datasets as a validation set. Figure \ref{fig:snrhatvsepoch} (right), instead, shows the SI-SNR computed using the simulated WHAMR! validation dataset.  


The training curve observed with the REAL-M dataset coupled with the proposed neural estimator looks pretty natural. As expected, it follows a standard logarithmic trend for all the models.  The performance improvement is larger in the first epochs, while a diminishing return is observed as long as training proceeds.
Moreover, the models evaluated on REAL-M with the neural estimator achieve the same performance ranking obtained on the validation set of WHAMR!.
In both cases, the best model is the SepFormer, followed by DPRNN and ConvTasNET.
This agreement is another indication of the reliability of the proposed estimator.

\vspace{-0.3cm}
\subsection{ASR based evaluation}
The REAL-M dataset also provides the text transcription of each signal in the mixtures.
Beyond using our blind SI-SNR estimator, it is thus possible to 
compute the Word-Error-Rate (WER) on the estimated sources. This metric can give another hint on the quality of the separator.

Table \ref{table:ASR} reports the average WER using different speech recognizers and separators.
In particular, we adopted a CRDNN model \cite{speechbrain} trained on the LibriSpeech (LS) dataset \cite{Panayotov_librispeech} and a Wav2Vec 2.0 (W2V2) \cite{baevski2020wav2vec} model finetuned on the Common Voice (CV)  English dataset \cite{commonvoice:2020}. Both models are implemented with SpeechBrain \cite{speechbrain}.


\vspace{-0.3cm}
\begin{table}[h!]
 \caption{
 Word-Error-Rates achieved on REAL-M
 with different speech recognizers and separators. The last column reports the estimated SI-SNR values.}
 \label{table:ASR}
 \centering
\resizebox{8.0cm}{!}{
\begin{tabular}{l|c|c|c}
\textbf{Separator} & \textbf{W2V2-CV $\downarrow$} & \textbf{CRDNN-LS $\downarrow$ } & $\widehat{\textbf{SNR}}$ $\uparrow$ \\
\hline \hline 
SF-WHAMR! & 60.7 & 77.3 & 2.88  \\ \hline 
DPRNN-WHAMR! & 64.9 & 78.4 & 2.43 \\ \hline 
CTN-WHAMR! & 72.8 & 85.8 & 1.59 \\ \hline 

\end{tabular}
}
\end{table}
We observe that the Wav2Vec 2.0 model outperforms the CRDNN one. This result further confirms the relative effectiveness of Wav2Vec 2.0 even in challenging acoustic conditions \cite{wav2vec}. The absolute speech recognition performance, however, is still poor. The high WER highlights one more time how challenging speech separation is under real-life conditions.
Also in this context, we observe the same performance ranking between the achieved WERs and the SI-SNR estimations of the proposed SI-SNR estimator.

\section{Conclusion}

In this work, we release REAL-M, a dataset for speech separation in real-life settings, obtained through crowd-sourcing. 
Moreover, we showed that a neural SI-SNR blind estimator can enable reliable evaluation of in-the-wild speech mixtures, for which the oracle target clean sources are unavailable. We extensively tested this approach, and we observed that the estimated SI-SNR values generally correlate well with the oracle SI-SNR values (on synthetic data) and with human assessments (on the real-life mixtures of REAL-M).

This study also highlights how challenging speech separation is on in-the-wild data. It is worth mentioning that speech separation in clean conditions (e.g., on WSJ0-2Mix) reaches 20 dB of SI-SNR with the best models for separation. The performance goes down to 8 dB when using simulated data with noise and reverberation (e.g., WHAMR!). It dramatically falls to 2.8 dB when using the real-life mixtures of REAL-M. Speech separation in real-life conditions is still an open problem, and we hope that our contribution will foster further research in this area.




We envision that, as future work, it is reasonable to explore the use of performance metrics other than SI-SNR, adapted from speech enhancement literature \cite{Pesq, stoi, Beerends_papersperceptual, hasqi, visqol}. It is also a natural next step to test approaches such as \cite{wisdom2020unsupervised} on the REAL-M dataset, as well as finetuning a separator using the neural SI-SNR estimator. 
\bibliographystyle{IEEEbib}
\bibliography{refs}

\end{document}